\documentclass[sigconf]{acmart}

\usepackage{multirow}
\usepackage{bm}
\usepackage{graphicx}
\usepackage{subcaption}

\AtBeginDocument{%
  }

\usepackage[english]{babel}

\usepackage{amsmath}
\usepackage{graphicx}

\title[Collaborative-Enhanced Prediction of Spending on Newly Downloaded Mobile Games \\ under Consumption Uncertainty]{Collaborative-Enhanced Prediction of Spending on Newly Downloaded Mobile Games under Consumption Uncertainty}

\author{Peijie Sun}
\affiliation{%
\institution{Quan Cheng Laboratory}
\institution{DCST, Tsinghua University}
  \city{Beijing}
  \country{China}}
\email{sun.hfut@gmail.com}

\author{Yifan Wang}
\affiliation{%
  \institution{DCST, Tsinghua University}
  \city{Beijing}
  \country{China}}
\email{yf-wang21@mails.tsinghua.edu.cn}

\author{Min Zhang}
\authornote{Corresponding author.}
\affiliation{%
  \institution{DCST, Tsinghua University}
  \city{Beijing}
  \country{China}}
\email{z-m@tsinghua.edu.cn}

\author{Chuhan Wu}
\affiliation{%
  \institution{Noah’s Ark Lab, Huawei}
  \city{Beijing}
  \country{China}}
\email{wuchuhan15@gmail.com}

\author{Yan Fang}
\affiliation{%
  \institution{DCST, Tsinghua University}
  \city{Beijing}
  \country{China}}
\email{fangy21@mails.tsinghua.edu.cn}

\author{Hong Zhu}
\affiliation{
  \institution{Consumer Cloud Service Interactive Media BU, Huawei}
  \city{Shenzhen}
  \country{China}}
\email{zhuhong8@huawei.com}

\author{Yuan Fang}
\affiliation{%
  \institution{Consumer Cloud Service Interactive Media BU, Huawei}
  \city{Shenzhen}
  \country{China}}
\email{frank.fy@huawei.com}

\author{Meng Wang}
\affiliation{%
  \institution{School of Computer Science and Information Engineering, Hefei University of Technology}
  \city{Hefei}
  \country{China}}
\email{eric.mengwang@gmail.com}

\renewcommand{\shortauthors}{Peijie Sun et al.}

\thanks{This work is supported by the Natural Science Foundation of China (Grant No.U21B2026, 72188101), the fellowship of China Postdoctoral Science Foundation (No.2022TQ0178), and Huawei (Huawei Innovation Research Program). We also thank MindSpore for the partial support of this work, which is a new deep learning computing framework.}

\copyrightyear{2024}
\acmYear{2024}
\setcopyright{rightsretained}
\acmConference[WWW '24 Companion]{Companion Proceedings of the ACM Web Conference 2024}{May 13--17, 2024}{Singapore, Singapore}
\acmBooktitle{Companion Proceedings of the ACM Web Conference 2024 (WWW '24 Companion), May 13--17, 2024, Singapore, Singapore}\acmDOI{10.1145/3589335.3648297}
\acmISBN{979-8-4007-0172-6/24/05}

\copyrightyear{2024}
\acmYear{2024}
\setcopyright{rightsretained}
\acmConference[WWW '24 Companion]{Companion Proceedings of the ACM Web Conference 2024}{May 13--17, 2024}{Singapore, Singapore}
\acmBooktitle{Companion Proceedings of the ACM Web Conference 2024 (WWW '24 Companion), May 13--17, 2024, Singapore, Singapore}\acmDOI{10.1145/3589335.3648297}
\acmISBN{979-8-4007-0172-6/24/05}

\makeatletter
\gdef\@copyrightpermission{
 \begin{minipage}{0.3\columnwidth}
  \href{https://creativecommons.org/licenses/by/4.0/}{\includegraphics[width=0.90\textwidth]{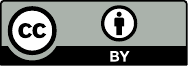}}
 \end{minipage}\hfill
 \begin{minipage}{0.7\columnwidth}
  \href{https://creativecommons.org/licenses/by/4.0/}{This work is licensed under a Creative Commons Attribution International 4.0 License.}
 \end{minipage}
 \vspace{5pt}
}
\makeatother

\begin{document}

\begin{abstract}
With the surge in mobile gaming, accurately predicting user spending on newly downloaded games has become paramount for maximizing revenue. However, the inherently unpredictable nature of user behavior poses significant challenges in this endeavor. To address this, we propose a robust model training and evaluation framework aimed at standardizing spending data to mitigate label variance and extremes, ensuring stability in the modeling process.
Within this framework, we introduce a collaborative-enhanced model designed to predict user game spending without relying on user IDs, thus ensuring user privacy and enabling seamless online training. Our model adopts a unique approach by separately representing user preferences and game features before merging them as input to the spending prediction module. Through rigorous experimentation, our approach demonstrates notable improvements over production models, achieving a remarkable \textbf{17.11}\% enhancement on offline data and an impressive \textbf{50.65}\% boost in an online A/B test.
In summary, our contributions underscore the importance of stable model training frameworks and the efficacy of collaborative-enhanced models in predicting user spending behavior in mobile gaming.
The code associated with this paper has also been released at the following link\footnote{\url{https://doi.org/10.5281/zenodo.10775846}}. 
\end{abstract}



\maketitle
\begin{figure}
  \centering
  \includegraphics[width=0.4\textwidth]{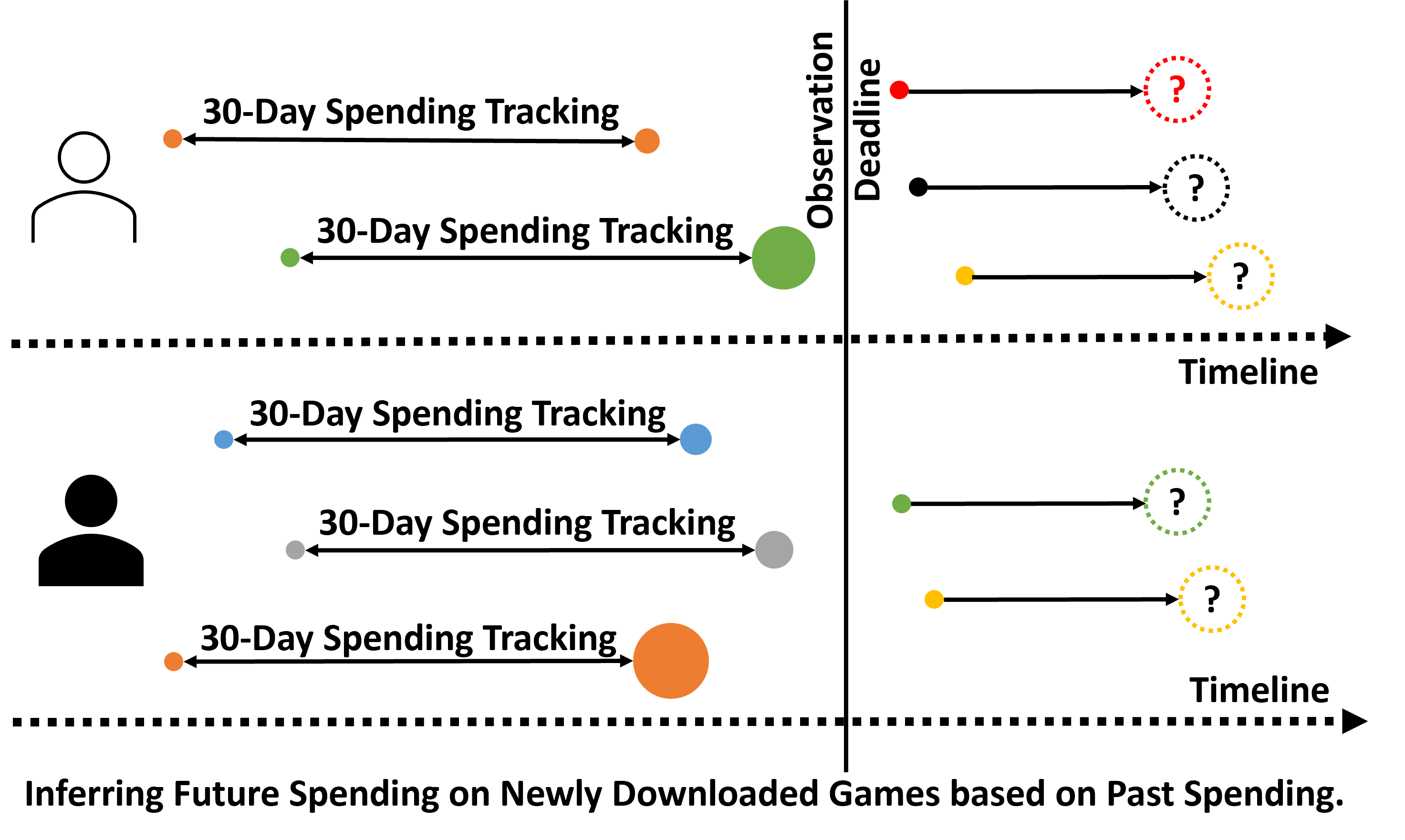}
  \caption{This illustration showcases our collected data. It includes observed spending behavior by users leading up to an observation deadline. Users' spending habits are tracked from the day they download the game, with a 30-day observation period to gauge their spending during the subsequent 30 days. Different colored circles represent different games, with larger circles indicating higher spending. Using this observed data, we aim to predict users' future spending within the next T days for newly downloaded games.}
  \label{fig:collected_data}
\end{figure}

\section{Introduction}
Online mobile games have become a part of many people's daily lives. Different game developers strive to earn money from mobile game players. 
However, convincing players to spend money depends not solely on the quality of the designed in-game goods. The game store's recommendation strategy is even more crucial. 
One potent way to achieve higher total revenue is by inferring how much the users may pay for newly downloaded games and formulating tailored recommendation strategies, such as the targeted Return On Ad Spend(tROAS)~\cite{arxiv2019ziln}. The task of inferring how much users pay for newly downloaded games can be treated as predicting the LifeTime Value (LTV) of these users~\cite{arxiv2019ziln}. The task of predicting LTV has been a topic of study for many years \cite{fader2005rfm, malthouse2005cltv, vanderveld2016engagement, li2022billion, win2020randomforest}. However, our approach represents a departure from these work, which studied how to predict a user's future expenditure based on past expenditures. Instead, our focus is on predicting the spending money of users for any new game they download. This method is a novel approach to maximizing revenue for game developers. 

One approach to this task is to convert LTV prediction into conversion rate(CVR) prediction, a technique employed by industry giants like Google, Tencent, and NetEase~\cite{arxiv2019ziln, arxiv2019ziln, xing2021tsur, zhao2023percltv}. Based on this approach, we need to prepare two things: data collection and model design. Given the continuous spending behavior of users and the varying life cycles of users in games, collecting the total amount spent by users is impractical. Instead, we use a 30-day observation window and consider the total spending amount within this period on newly downloaded games as the user's LTV. In other words, the LTV would be treated as the label to train the model which is designed for the CVR task. By minimizing the LTV label and the prediction of the CVR model, we can treat the final learned CVR model as an LTV prediction model. The illustration of the dataset we collected can refer to Figure~\ref{fig:collected_data}.
Our current production environment utilizes a classical cross-feature CVR model as the backbone model. This model inputs the users' historical behaviors, such as downloaded and paid game apps, and games' profiles like game ID and other related metadata. The model aims to predict the LTV of the user-game pair, and we optimize it by minimizing the difference between predicted spending and real observed spending. 

\begin{figure}[htb]
    \centering
    \includegraphics[width=0.49\columnwidth]{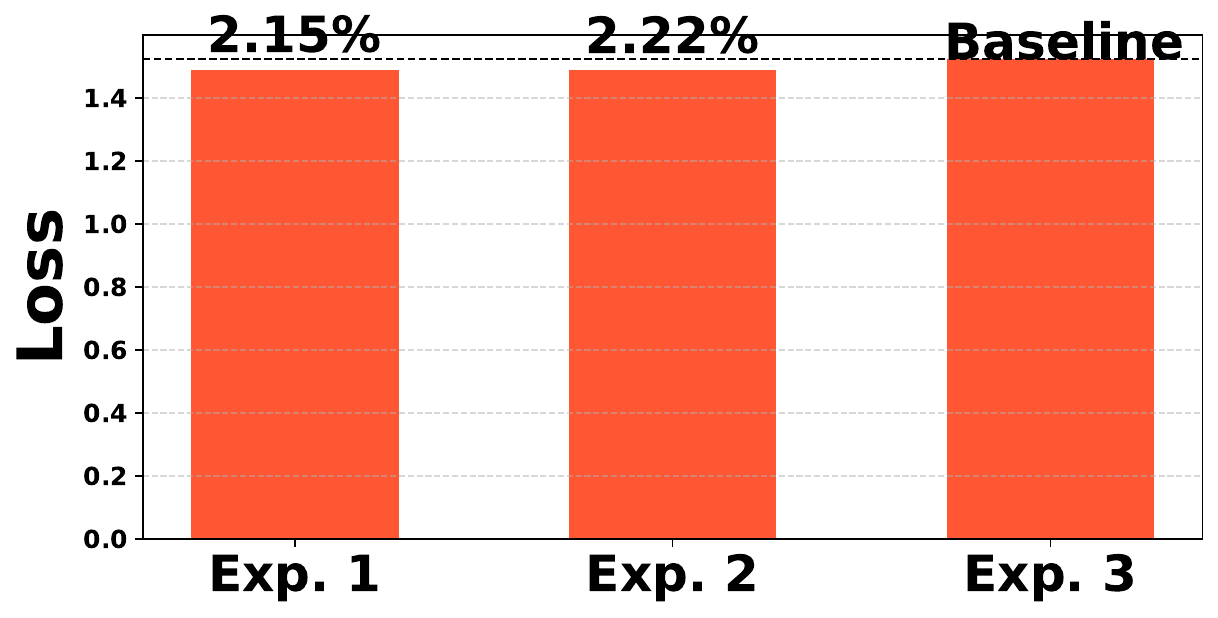} 
    \hfill
    \includegraphics[width=0.49\columnwidth]{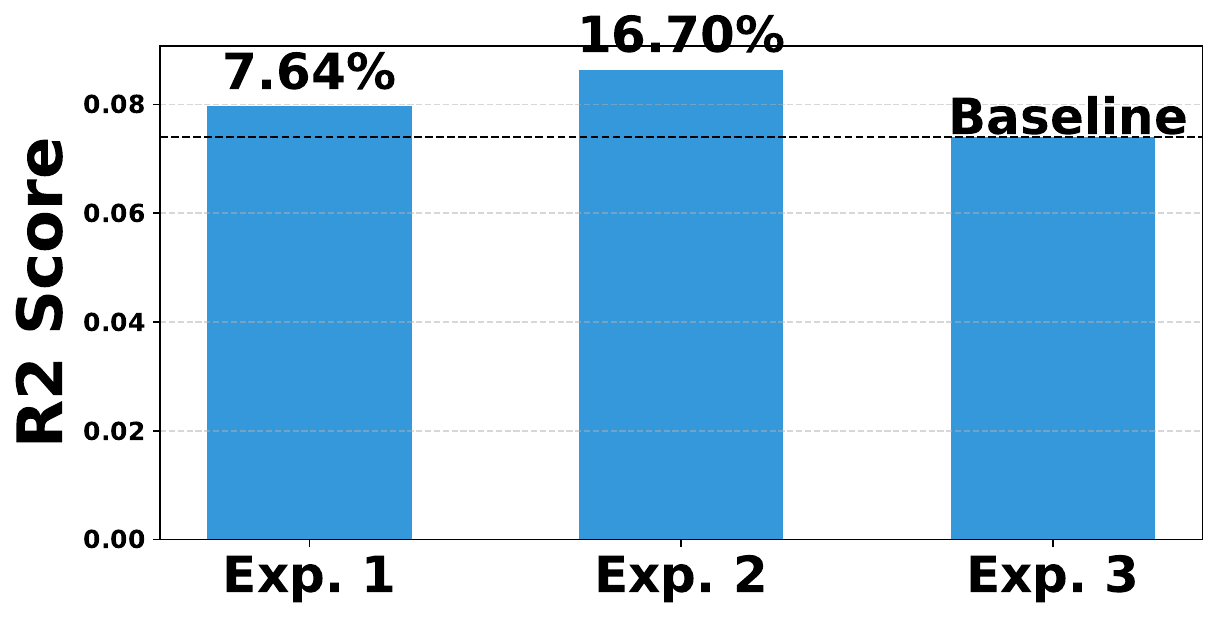} 
    \caption{Repeated three production experiments, labeled Exp. 1, Exp. 2, and Exp. 3, were conducted using a private dataset to evaluate the spending money prediction capacity. RMSE-based loss is used as objective loss, and R2 Score~\cite{Wikipedia_2023_r2score} is used as model effectiveness evaluation. The worst-performing model is marked with the baseline. The numbers above each bar indicate improvements compared to the baseline. For example, the Exp. 1 bar, represents a 2.15\% improvement in training loss and a 7.64\% improvement in the R2 Score over the worst-performing experiment.} 
    \label{fig:my_label}
\end{figure}

Upon reviewing the spending prediction task, we argue that using only the cross-feature model as the CVR model is somewhat not enough, as this type of model does not explicitly consider the collaborative relationship~\cite{mnih2007pmf, rendle2012bpr, www2017neumf, sigir2020lightgcn, sigir2021sgl, tkde2023neigborhood, cikm2021simplex, sun2023nescl} between users.
However, collaborative signals are very important for predicting spending amounts. Because a user is more likely to spend more money on her satisfied product. Utilizing the collaborative signals can help the model better find the product that meets the target users' preferences. However, giving collaborative capability to the current cross-feature production model is not a simple task.
This is mainly due to the unique characteristics of our dataset.
Our data is characterized by many zero values, high label variance in non-zero values, and extremely high values, which brings significant challenges for consistently training and evaluating the effectiveness of the designed model. 

The dataset's characteristics are summarized in Table~\ref{tab:stat}. To illustrate, the dataset contains many zero values, but this doesn't mean users always spend nothing on games. Inaccurate consumption tags can harm model optimization and performance. High label variance and extreme values disrupt optimization and training stability. Relying solely on labels for model accuracy is problematic due to unpredictable user spending patterns.
As a result, we risk creating a model effective only on the validation dataset, lacking robustness for broader use. Though data characteristics also affect existing production models, experiments show that different models can perform well under the production strategy despite fluctuations. We suppose that the possible reason is adequate data can help mitigate uncertainties from the collected dataset.
Nevertheless, operating within the production environment continues to pose challenges for exploring new design models. For instance, in Figure~\ref{fig:my_label}, three experiments were conducted with the production model, revealing that while training loss remains stable, the overall performance, as measured by the R2 Score metric, varies significantly. Compared to the worst experiment, training loss only improves by 2.15\% and 2.22\%, while the R2 Score shows substantial improvements of 7.64\% and 16.70\%.

Our study has two main goals. First, we propose a framework to enhance model training and evaluation stability by addressing data characteristics. Second, we investigate collaborative signal modeling within this framework to develop a collaborative-enhanced solution. 
In our framework, we adhere to two key principles for effective model deployment. Firstly, we use spending data as labels during model training, employing a regression loss function. Secondly, we prioritize model stability and evaluation certainty by standardizing user spending mitigating label variance and extreme values. 
Compared with the value-based metric(minimizing the difference between the predicted and real spending money), we utilize the leave-one-out ranking metric~\cite{www2017neumf} to evaluate the trained model. When evaluating the model's effectiveness, we only evaluate whether the model can predict which game a user downloaded relatively accurately, without evaluating whether the user pays, and how much, because user downloading behavior is more easily learned in the dataset than the uncertain paying behavior. We also name this task the downloaded game prediction task. The experimental results also show that our proposed framework can bring out the capabilities of different models and that models validated as effective in our framework are also considered effective in predicting the amount of money a user will spend in the production environment, i.e., in the spending money prediction task. 

While designing the collaborative-enhanced model, we address two key questions:
First, how can we model the collaborative signal without using user IDs, which are typically essential for collaborative filtering, in order to protect user privacy? To tackle this, we represent user preferences using their historical game app downloads. We employ a multi-layer perceptron (MLP) with download history as input, obtaining the collaborative representation by linearly combining user preference and game data.
Second, how can we integrate this collaborative signal into the existing production model? We use the same behavior data for both the deployed cross-feature model and collaborative signal modeling. Since the online model doesn't explicitly capture the collaborative signal, we propose enhancing spending prediction by concatenating the collaborative feature with the cross-feature model output and inputting these concatenated features into the subsequent spending prediction module.
With no need for user IDs during model training and shared historical download game IDs across data samples, the model can be updated based on streaming data, making it adaptable for online training environments. Moreover, this approach ensures user privacy protection during model training. 

In summary, the contributions of our work can be summarized as:
1. Given the uncertainty surrounding users' consumption habits in terms of time and money, we have developed a framework designed to expedite exploring the collaborative-enhanced model. This framework is precisely engineered to ensure stable model training and evaluation processes. 

2. We have introduced a collaborative-enhanced model that prioritizes user privacy by not relying on user IDs. This approach allows for model training within a data streaming environment, as user IDs are unnecessary. 

3. Our collaborative-enhanced model has proven effective in two settings: First, on the offline business dataset, it outperformed the production model by \textbf{17.11}\% in the 30-day spending money prediction task.
Second, in the online A/B test, it achieved a \textbf{50.65}\% improvement in user payment revenue over two weeks compared to the production model.


\section{Dataset Statistics}

\begin{table}[]
\small
\caption{Statistics Of The Collected Private Dataset.}
\label{tab:stat}
\begin{tabular}{|ll|ll|}
\hline
\multicolumn{2}{|c|}{Consumption}                 & \multicolumn{2}{c|}{\begin{tabular}[c]{@{}c@{}}Download Game Apps \\ in Two Weeks\end{tabular}} \\ \hline
\multicolumn{1}{|l|}{Min Cost}       & 0.01       & \multicolumn{1}{l|}{Min Length}                                     & 1                         \\ \hline
\multicolumn{1}{|l|}{Max Cost}       & 421,387    & \multicolumn{1}{l|}{Max Length}                                     & 10                        \\ \hline
\multicolumn{1}{|l|}{Avg Cost}       & 227.15622  & \multicolumn{1}{l|}{Avg Length}                                     & 3.78                      \\ \hline
\multicolumn{1}{|l|}{Std}            & 2130.0593  & \multicolumn{1}{l|}{Std}                                            & 3.41                      \\ \hline
\multicolumn{1}{|l|}{Medium}         & 11.0       & \multicolumn{1}{l|}{Medium}                                         & 2                         \\ \hline
\multicolumn{1}{|l|}{\# of Non-zero} & 623,075    & \multicolumn{1}{l|}{\# of All Download Apps}                         & 7,507                     \\ \hline
\multicolumn{1}{|l|}{\# of Zero}     & 29,076,679 & \multicolumn{1}{l|}{\# of Filtered Paid Games}                      & 432                       \\ \hline
\end{tabular}
\end{table}

Our research aims to understand how to predict users' future spending on newly downloaded games. To collect the dataset, we adopted a straightforward approach. We initiated a T-day observation period for each user and their newly downloaded game. During this time, we closely monitored the user's spending on the game within the T-day window since the game's download date. The data collection spanned from August 1, 2022, to August 31, 2022.
Let's take August 7, 2022, as an example to provide a clearer illustration of the collected data. On this date, we gathered information regarding how much users spent on games they had downloaded on July 7, 2022. This extended observation period was essential to accumulate a sufficient volume of user behavior data for our analysis.
The statistics of the collected data can be referred to Table~\ref{tab:stat}.
Given our primary focus on paid games, we implemented a filter to exclude games with spending below a certain threshold. We cannot disclose the specific threshold used to safeguard company proprietary information. This filtering process accounts for the disparity between the total number of downloaded games and the number of paid games.
The downloaded game applications are crucial for modeling user preferences in our subsequent work.

From the collected data, we have the following observations:
\textbf{Prevalence of Zero Values}:
The dataset prominently displays a substantial prevalence of zero values, encompassing an impressive count of 29,076,679 zero samples, in stark contrast to the 623,075 non-zero samples. This notable contrast implies a significant presence of users who did not partake in any spending activity. It's important to note that due to the limited observation time window, these zero values do not necessarily indicate that users consistently spent nothing on the corresponding games.
\textbf{High Variance Among Non-Zero Samples}:
While the average consumption cost (Avg Cost) appears relatively low at 227.16, the standard deviation (Std) stands at a remarkably high 2130.06. This disparity implies substantial variability within non-zero consumption costs, with some values significantly deviating from the average. The resulting high standard deviation underscores the substantial variance in the dataset. 
\textbf{Extreme Values}:
Notably, the maximum consumption cost (Max Cost) reaches an extreme value of 421,387, indicating the existence of outliers with exceptionally high spending. Similarly, the elevated standard deviation (Std) indicates significant data points deviating from the mean. These findings collectively highlight the presence of outliers, potentially attributed to users with exceptionally high expenditures. 

Directly using these significantly high-variance consumption amounts as labels would introduce varying degrees of gradient values to the trainable parameters, potentially severely disrupting their normal learning process. Additionally, during the model evaluation phase, predicting user behaviors concerning these zero and extreme values could be hindered by the uncertainty surrounding user spending time and amounts. This uncertainty might result in models selected based on a specific validation set lacking the required level of generalization.
\section{Proposed Framework And Collaborative-Enhanced Model}

\begin{figure}[ht]
    \centering
    \begin{subfigure}{0.40\columnwidth}
        \centering
        \includegraphics[width=\linewidth]{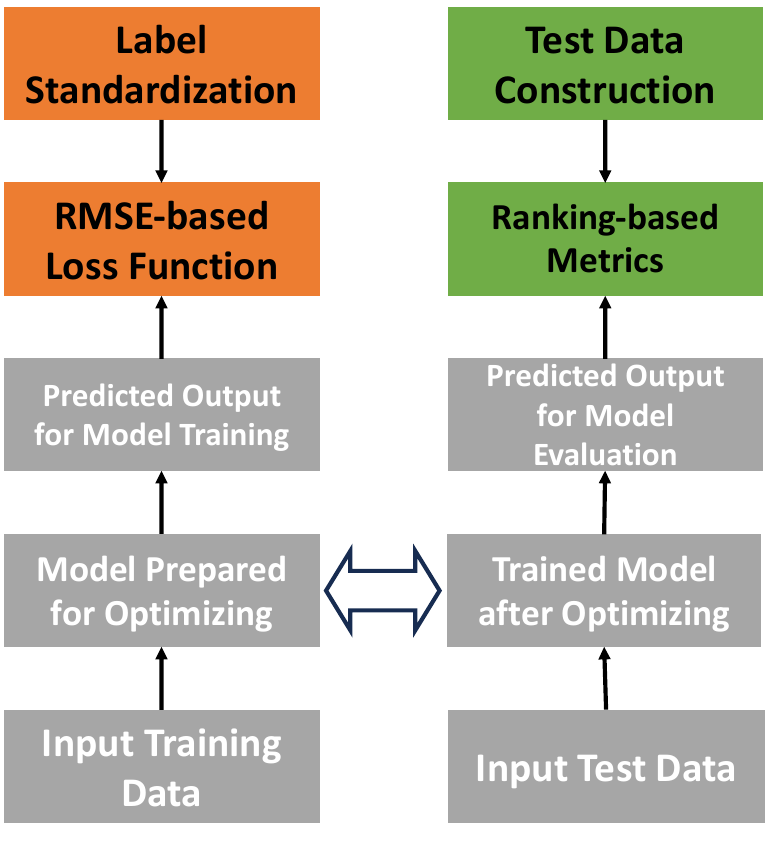}
        \caption{Our Proposed Framework}
        \label{fig:hr10}
    \end{subfigure}
    \hfill 
    \begin{subfigure}{0.55\columnwidth}
        \centering
        \includegraphics[width=\linewidth]{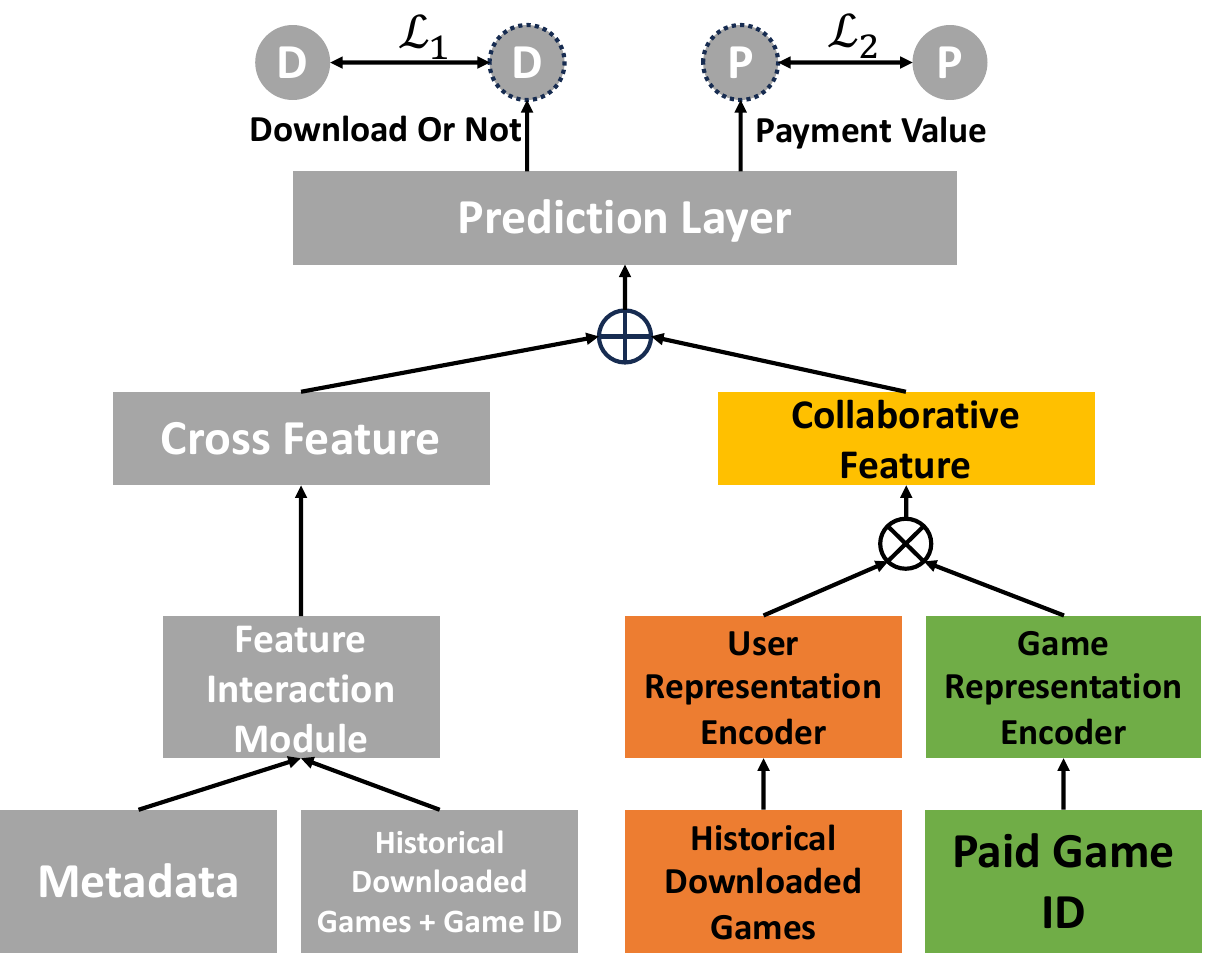}
        \caption{Our Proposed Collaborative-Enhanced Production Model}
        \label{fig:ndcg10}
    \end{subfigure}
    \caption{The Comprehensive Architecture of Our Proposed Framework and Model. The modules colored other than gray represent the new modules we have introduced.}
    \label{fig:metrics}
\end{figure}

In this section, we will first introduce the proposed framework to accelerate the exploration of newly designed models and the collaborative-enhanced model we have developed. Initially, we will introduce the notations employed in our proposed model, which can be found in Table~\ref{tab:notation}.

\begin{table}
    \centering
    \footnotesize
    \caption{Notation Table.}
    \begin{tabular}{|c|c|}
    \hline
       $\mathcal{U}=(u_1, u_2, ..., u_{|\mathcal{U}|})$ & User set \\
        \hline
        $\mathcal{P}=(p_1, p_2, ..., p_{|\mathcal{P}|})$ & Paid Game set \\
        \hline
        $\mathcal{D}=(d_1, d_2, ..., d_{|\mathcal{D}|})$ & Downloaded Game set \\
        \hline
        $s_{up}$ & Spent money from user-game pair $(u, p)$\\
         \hline
         $t_u^{180}$ & Total spending of user $u$ in the past 180 days\\ 
         \hline
         $f_u^{180}$ & Total payment times of user $u$ in the past 180 days\\ 
         \hline
        $h_u^{d}$ & Downloaded game apps list for the user $u$\\
         \hline
    \end{tabular}
    \label{tab:notation}
\end{table}

\subsection{Proposed Framework}
Our proposed framework is designed to address the unique characteristics of the collected datasets, including numerous zero values, high variance in non-zero labels, and exceptionally high values. We have enhanced both the training and evaluation strategies within the production environment. The advanced modules we have introduced are depicted as non-gray modules in Figure~\ref{fig:my_label}(a).
In the left section of Figure~\ref{fig:my_label}(a), we have introduced label standardization and an RMSE-based loss function to train the newly designed model. Label standardization mitigates the adverse effects of high label variance in non-zero and highly high values during model optimization. The RMSE-based loss function assesses whether the newly designed model can capture patterns under the same labels and training conditions as those in the production environment.
In the right section of Figure~\ref{fig:my_label}(a), we employ a ranking-based method to evaluate the model's performance. This method verifies whether the proposed model can predict users' downloaded games from a mix of positive and negative examples. Such information is obtainable from the training data and helps reduce the impact of uncertain user spending patterns and detailed monetary data. 

\subsubsection{Label Standardization}
To standardize the spending labels, we aim to uncover their underlying information. We argue that the absolute amount spent does not necessarily reflect a user's preference for a game. Some low-cost games may also hold value for users. User spending habits and cumulative spending for the game can influence larger or smaller spending amounts. The standardized label serves to assess the game's value to the user. Treating this value as the label is reasonable, as it helps the model identify valuable games for the user. This information is crucial for predicting game downloads and estimating potential spending, as users tend to spend more on games they find valuable.

Specifically, we collect the users' total spending $t_u^{180}$ and payment times $f^{180}_u$ in the past 180 days before the observation time and cumulative spending for the games $\mathcal{S}_p$ to standardize the spending money labels. Given user $u$ and paid game $p$, we utilize the following equation to standardize the collected spending money $s_{up}$:
\begin{equation}
\tilde{s}_{up} = f(t_u^{180}, f^{180}_u, \mathcal{S}_p, s_{up}).
\end{equation}
The function $f$ denotes the standardization function, and the notation $\tilde{s}_{up}$ denotes the standardization cost money. 

We have adopted a strategy that involves first calculating the standardization values separately from the game's and user's perspectives and then combining them to form the final standardization value. For game-sided standardization, we compute the corresponding standardized spending value as follows:

\begin{equation}
    \tilde{s}_p = \frac{s_{up}-AVG(\mathcal{S}_p)}{STD(\mathcal{S}_p)},
    \label{eq:game-sided}
\end{equation}

Where $AVG(\cdot)$ represents the average function, used to calculate the mean of all historical spending for game $p$, and $STD(\cdot)$ represents the standard deviation of all historical spending for game $p$.

From the user's perspective, we employ the following equation to calculate the user-sided standardization value:

\begin{equation}
    \tilde{s}_u =  \frac{s_{up}}{t_u^{180}/ f^{180}_u}.
    \label{eq:user-sided}
\end{equation}
During the standardization process, our goal is to rescale the consumption data of different users to a similar scale to eliminate scale differences among them.

Due to the distinct scales between the values obtained by users and the standardized label values obtained by games, we combine all the standardized labels of users and games to normalize their respective standardized labels. Subsequently, we sum and aggregate them to obtain the final standardized labels, represented as $\tilde{s}_{up}=0.5\tilde{s}_p+0.5\tilde{s}_u$.

\subsubsection{Model Optimization}
In the model optimization stage, we use the MSE loss to optimize the backbone model within the framework. The objective loss can be calculated with:
\begin{equation}
    \mathcal{L}_{MSE} = \frac{\sum_{(u,p)\in\mathcal{O}}(\tilde{s}_{up}-\hat{s}_{up})^2}{|\mathcal{O}|},
\end{equation}
Where $\mathcal{O}$ denotes all observed user-game consumption records. And $\hat{s}_{up}$ denotes the predicted output of the backbone model. 

\subsubsection{Model Evaluation}
When assessing the effectiveness of the trained model, our objective is to determine whether the model can predict which games users might download from a mixture of positive and negative games. This task relies on information that can be captured during the training process. Specifically, for each test sample, we randomly selected 100 games from all paid games to serve as negative examples. This strategy is also employed in BPR~\cite{rendle2012bpr} and NeuMF~\cite{www2017neumf}.
We evaluate the model's performance using ranking-based metrics, including NDCG@K and HR@K~\cite{www2017neumf}. HR@K indicates whether the user's interacted game ranks in the top-K game list. On the other hand, NDCG@K assesses whether the interacted game ranks first. Generally, the higher a game is ranked, the greater the value of the corresponding metric, indicating superior performance by the trained model.

\subsection{Collaborative-Enhanced Model}
This section will introduce how to model the collaborative signal and enhance the production model. The updates to the proposed model can be observed in the non-gray modules of Figure~\ref{fig:my_label}(b). The gray section represents the simplified structure of our online production model. To safeguard the company's proprietary information, we won't delve into it extensively in this paper.
Since the production model utilizes the same data, such as users' historical download game apps and paid game IDs, we argue that it can capture implicit collaborative signals. Our proposed explicit collaborative signals should prove valuable in enhancing the model's predictive power for users' spending. As for why we don't replace it with our proposed collaborative signal, it's primarily because the production model employs a complex mapping function to transform features into the final spending amount. The input features should include metadata, such as users' historical spending data. However, this paper doesn't delve into how to combine the historical spending input with our proposed collaborative signals. We treat the collaborative signal as helpful and rely on the existing production environment to harness its value.

First, we will introduce how to model the collaborative signal without using user IDs to protect users' privacy. The goal of user representation is to identify users with similar preferences, and for this purpose, we can leverage users' historical behaviors to represent their embeddings. Ultimately, we choose to utilize users' historical downloaded games as they are easier to collect and rich in information.

Given a user \( u \)'s downloaded game list \( h^u_d = \{ d_1, d_2, \ldots, d_{10} \} \), we can obtain the user's preference representation as follows:

\begin{equation}
\mathbf{v}_u = MLP\left(\begin{bmatrix}
    \mathbf{e}_1 & 
    \mathbf{e}_2 &
    \dots &
    \mathbf{e}_{|h^u_d|}
    \end{bmatrix}\right),
\end{equation}
where \(\mathbf{e}_{|h^u_d|}\) denotes the embedding of the \(|h^u_d|\)-th downloaded game in user \(u\)'s download history.

The collaborative signal can be modeled using the following equation:

\begin{equation}
\mathbf{v}_{up} = \mathbf{v}_u * \mathbf{v}_p,
\end{equation}
where \(\mathbf{v}_p\) represents the embedding of the paid game.

Next, we use the concatenate operation to combine the collaborative signal \(\mathbf{v}_{up}\) with the existing production model:

\begin{equation}
\mathbf{v}_{concat} = \begin{bmatrix}
    \mathbf{v}_{up} & \mathbf{v}^{fi}_{up}
    \end{bmatrix}
\end{equation}

The final spending amount can be predicted using the complex mapping function currently employed in the production environment:

\begin{equation}
    \hat{s}_{up} = f(\mathbf{v}_{concat})
\end{equation}

\section{Experiments}
In this section, we aim to validate the effectiveness of our proposed framework and collaborative-enhanced model. The experiments can be divided into two main parts.
The first part is designed to confirm the viability and stability of our proposed framework in the downloaded game prediction task. We seek to verify whether our framework can aid the designed model in uncovering hidden patterns within the dataset. Additionally, we aim to assess whether models with varying capabilities can perform at their full potential. 
In the second part, we evaluate the effectiveness of our proposed collaborative-enhanced model in both offline and online production environments in the spending money prediction task. 

\subsection{Experimental Settings}
The latent factor dimension size is configured at 8, while the learning rate is set to 1e-5. Subsequently, we employ a 4-layer MLP with layer sizes [8, 16, 32, 8] to model users' preference representations. The Adam optimizer guides the optimization process, which updates the trainable parameters. For training, we utilize the PyTorch deep learning framework on a GPU server equipped with a 32GB GPU card, 64GB of memory, and a 32-core 2.3 GHz CPU.

\subsection{Effectiveness of Our Proposed Framework}

\subsubsection{Framework Availability}
\begin{figure}[]  
  \centering
  \includegraphics[width=0.95\linewidth]{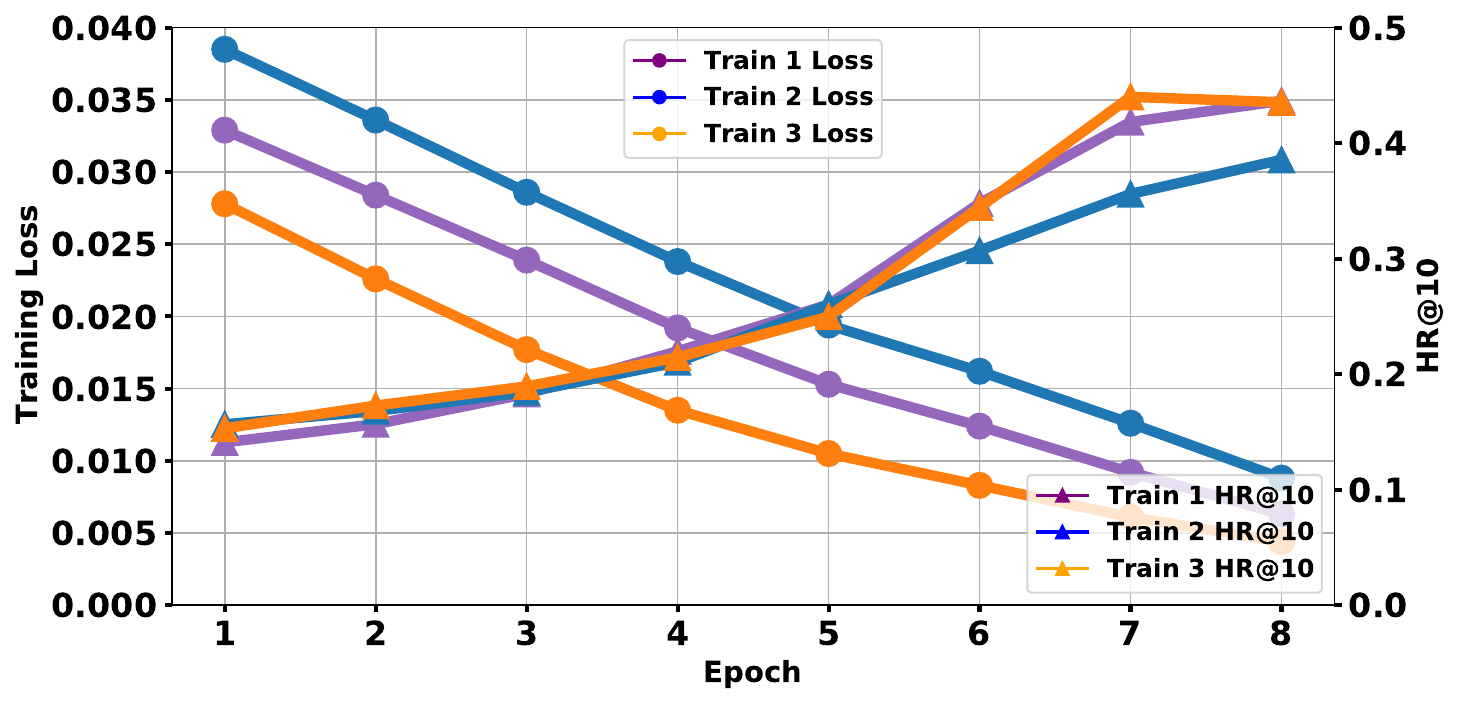}  
  \caption{Training and Evaluation Process Curve}
  \label{fig:training-results}
\end{figure}

In this study, we conducted multiple training sessions to assess the effectiveness of our recommendation framework. Our primary focus was on whether the model could uncover hidden patterns within the dataset. We closely monitored the model's performance throughout the training process. To ensure the stability of our framework, we repeated the experiment three times and generated training and evaluation curves depicted in Figure~\ref{fig:training-results}.
Consistently, we observed that performance metrics, such as HR@10, displayed consistent and notable improvements across all training sessions. These findings highlight the model's capacity to enhance its recommendation capabilities as it undergoes training. 

\subsubsection{Framework Stability - Different Backbone Models.}

\begin{table}[]
\footnotesize
\caption{Performance of Recommendation Models on HR@K and NDCG@K.}
\label{tab:over_our_frame}
\begin{tabular}{|c|ccc|ccc|}
\hline
\multirow{2}{*}{Method} & \multicolumn{3}{c|}{HR@K$\uparrow$}                                          & \multicolumn{3}{c|}{NDCG@K$\uparrow$}                                        \\ \cline{2-7} 
                        & \multicolumn{1}{c|}{K=1}    & \multicolumn{1}{c|}{K=5}    & K=10   & \multicolumn{1}{c|}{K=1}    & \multicolumn{1}{c|}{K=5}    & K=10   \\ \hline
MF~\cite{mnih2007pmf}                     & \multicolumn{1}{c|}{0.0185} & \multicolumn{1}{c|}{0.1275} & 0.271  & \multicolumn{1}{c|}{0.0094} & \multicolumn{1}{c|}{0.0687} & 0.1186 \\ \hline
FM~\cite{idcm2010fm}                      & \multicolumn{1}{c|}{0.0415} & \multicolumn{1}{c|}{0.1653} & 0.3834 & \multicolumn{1}{c|}{0.0262} & \multicolumn{1}{c|}{0.0893} & 0.1207 \\ \hline
Wide \& Deep~\cite{cheng2016widedeep}            & \multicolumn{1}{c|}{0.0495} & \multicolumn{1}{c|}{0.2839} & 0.4542 & \multicolumn{1}{c|}{0.0306} & \multicolumn{1}{c|}{0.1807} & 0.2386 \\ \hline
DCNV2~\cite{www2021dcnv2}                   & \multicolumn{1}{c|}{0.0702} & \multicolumn{1}{c|}{0.3322} & 0.5024 & \multicolumn{1}{c|}{0.0403} & \multicolumn{1}{c|}{0.1990}  & 0.2546 \\ \hline
\textbf{OURS}                    & \multicolumn{1}{c|}{0.0830}  & \multicolumn{1}{c|}{0.3687} & 0.5477 & \multicolumn{1}{c|}{0.0524} & \multicolumn{1}{c|}{0.2266} & 0.2841 \\ \hline
\end{tabular}
\end{table}

In this section, to assess the stability of our proposed framework, we aim to determine its effectiveness in enabling various models to perform at their full potential. We have chosen a classical collaborative-based model, mf~\cite{mnih2007pmf}, and a classical cross-feature model, fm~\cite{idcm2010fm}. Additionally, we have included two competitive cross-feature models,~\cite{cheng2016widedeep} and \cite{www2021dcnv2}. The results can be found in Table\ref{tab:over_our_frame}. Our proposed model is denoted as \textbf{OURS}, representing an enhanced version of DCNV2 that explicitly models collaborative signals.

The results consistently revealed a linear improvement in performance from MF to \textbf{OURS}, indicating an advancement in modeling techniques. Particularly, \textbf{OURS} exhibited superior performance across all K values in both metrics, underscoring its effectiveness in prioritizing user-downloaded games within candidate game lists. An interesting trend noted was the increase in HR@K and NDCG@K scores with higher K values for all models, reflecting the broadening opportunity to capture user preferences. Additionally, the performance gap between models varied with different K values, with significant disparities becoming evident at higher Ks. These insights collectively demonstrate the robustness of 'OURS' in enhancing recommendation relevance and precision compared to other models. 

Overall, the results indicate that our proposed framework assists various models in realizing their full potential and serves as inspiration. It highlights that the collaborative signal modeling method and the combination approach can be effectively applied in a production environment.

\subsubsection{Framework Stability - Label Standardization}
\begin{figure}[]
    \centering
    \begin{subfigure}{0.49\columnwidth}
        \centering
        \includegraphics[width=\linewidth]{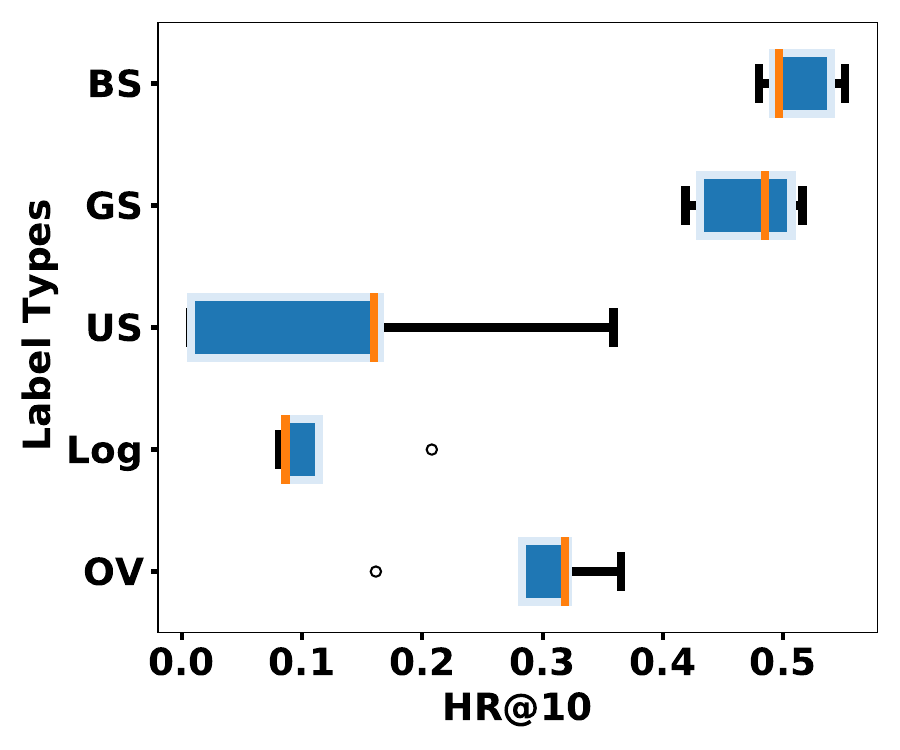}
        \caption{HR@10 Performance}
        \label{fig:hr10}
    \end{subfigure}
    \begin{subfigure}{0.49\columnwidth}
        \centering
        \includegraphics[width=\linewidth]{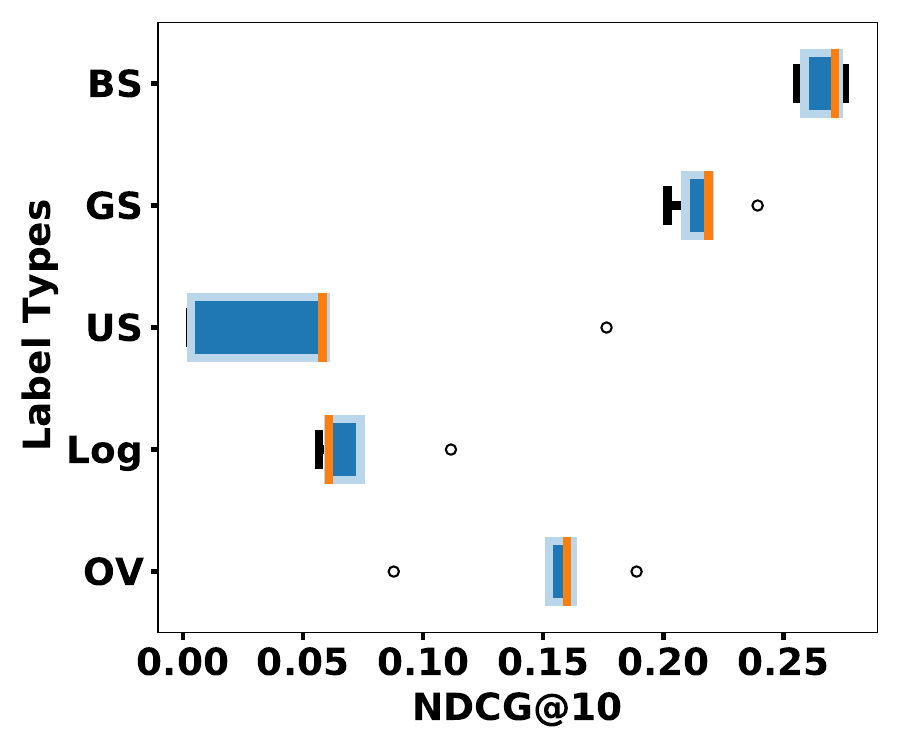}
        \caption{NDCG@10 Performance}
        \label{fig:ndcg10}
    \end{subfigure}
    \caption{Performance of our proposed model under different label standardization types. BS denotes both-sided, GS denotes game-sided, US denotes user-sided, Log denotes logarithmic, and OV denotes origin values.}
    \label{fig:lab_stand}
\end{figure}

In this section, we aim to verify the effectiveness of our proposed label standardization strategies. The five label standardization methods are Original Values(OV), Logarithmic(Log), User-sided(US), Game-sided(GS), and Both-sided(BS). 
OV denotes we treat original spending values as labels directly. Log denotes we utilize the logarithmic value of the spending values as labels. US denotes we only treat the user-sided standardization label as a supervision signal, such as Equation~\ref{eq:user-sided}. Similarly, the GS denotes we only treat the user-sided standardization label as a supervision signal, such as Equation~\ref{eq:game-sided}. 
From Figure~\ref{fig:lab_stand}, we find Original Values show a moderate, steady performance, underscoring the raw data's inherent predictive strength. Conversely, the Logarithmic approach consistently underperforms, suggesting it might not fit this data set best. While effective in certain scenarios, the user-sided method falls short in terms of its overall performance. In a comprehensive evaluation, it lags behind other approaches.
The Game-sided strategy scores high across all labels, making it a generally effective option. However, the star performer is the Both-sided approach, excelling in every category. This method's success is likely due to its holistic consideration of user and game factors. This insight highlights the Both-sided method as the most effective overall. Furthermore, the compactness of the box plot corresponding to both sides indicates the stability of the results achieved through this standardized approach.

\subsubsection{Framework Stability - Evaluation}

\begin{table}[]
\caption{Evaluation of Model Generalization in Different Environments. CoV(Coefficient of Variation)~\cite{Wikipedia_2023_cov} is a measure of data variability. CoV$\downarrow$($\uparrow$): a lower(higher) value means more consistent(greater variability).}
\label{tab:gener}
\footnotesize
\begin{tabular}{|c|c|c|c|c|}
\hline
Environment                                                                & Metric  & Mean    & Std     & CoV$\downarrow$     \\ \hline
Production                                                                 & R2Score & 0.02907 & 0.01389 & 0.47785 \\ \hline
\multirow{2}{*}{\begin{tabular}[c]{@{}c@{}}Our  \\ Framework\end{tabular}} & HR@10   & 0.42701 & 0.01942 & 0.04547 \\ \cline{2-5} 
                                                                           & NDCG@10 & 0.23274 & 0.01128 & 0.04848 \\ \hline
\end{tabular}
\end{table}

This section aims to demonstrate that our proposed framework allows for a fairer evaluation of our learned model's generalization capabilities. Our framework effectively measures the model's generalization because it assesses whether our proposed model can accurately identify games users have downloaded from a mix of positive and negative games. In contrast, the production model's task is to predict users' future spending, which can be uncertain and unstable, leading to variable model performance from one day to another.
To validate this, we conducted experiments using user spending data collected for training from August 1, 2022, to August 24, 2022. Subsequently, we evaluated the model's performance daily from August 25, 2022, to August 31, 2022, generating daily performance metrics.
We exclusively assessed the learned model using the R2Score metric in the production environment. However, within our proposed framework, we evaluated its performance using HR@10 and NDCG@10 metrics. We used the same backbone model for both evaluation approaches to ensure a fair comparison.
To simplify the presentation of results and account for metric scale differences, we provided the mean, standard deviation, and coefficient of variation (CoV)~\cite{Wikipedia_2023_cov} for the performance metrics over the seven days. 
Please refer to the Table~\ref{tab:gener} for detailed results.

This table shows that the Coefficient of Variance (CoV) value within our framework is notably lower compared to the production environment. 
This observation supports the idea that our proposed framework can effectively identify models with stable performance in the downloaded game recommendation task.
Furthermore, models that exhibit superior performance within our proposed framework will likely excel in the user spending prediction task. Our evaluation approach remains unaffected by the uncertainty surrounding spending time and amounts. Moreover, the model's training is primarily based on spending money. Consequently, the capabilities demonstrated by models within our framework will likely seamlessly transition into an online environment.

\subsection{Effectiveness of Our Proposed Collaborative-Enhanced Model}

\subsubsection{Evaluation on Offline Business Dataset}

\begin{table}[]
\footnotesize
\caption{Overall Performance on Private Dataset (Offline Testing). * indicates significant superiority over the top comparison (t-test, p<0.05). $\uparrow$($\downarrow$): the larger(the lower) the better.}
\label{tab:over_offline}
\begin{tabular}{|c|ccc|cc|}
\hline
\multirow{2}{*}{Method} & \multicolumn{3}{c|}{All Samples}                                      & \multicolumn{2}{c|}{\begin{tabular}[c]{@{}c@{}}Samples with \\ Payment   \textgreater 0\end{tabular}} \\ \cline{2-6} 
                        & \multicolumn{1}{c|}{RMSE$\downarrow$}   & \multicolumn{1}{c|}{R2 Score$\uparrow$} & AUC$\uparrow$    & \multicolumn{1}{c|}{RMSE$\downarrow$}                                  & R2 Score$\uparrow$                                \\ \hline
Linear                  & \multicolumn{1}{c|}{247.16} & \multicolumn{1}{c|}{0.01958}   & 0.7517 & \multicolumn{1}{c|}{1770.4}                                & 0.02588                                  \\ \hline
MLP                     & \multicolumn{1}{c|}{248.11} & \multicolumn{1}{c|}{0.01209}   & 0.6549 & \multicolumn{1}{c|}{1789.0}                                  & 0.00538                                  \\ \hline
Random Forest~\cite{win2020randomforest}                      & \multicolumn{1}{c|}{248.87} & \multicolumn{1}{c|}{0.00599}   & 0.7317 & \multicolumn{1}{c|}{1803.0}                                  & 0.01035                                  \\ \hline
XGBoost~\cite{chen2016xgboost}                 & \multicolumn{1}{c|}{246.40}  & \multicolumn{1}{c|}{0.02565}   & 0.8336 & \multicolumn{1}{c|}{1786.3}                                & 0.00835                                  \\ \hline
ZILN~\cite{arxiv2019ziln}                    & \multicolumn{1}{c|}{242.56} & \multicolumn{1}{c|}{0.05576}   & 0.8601 & \multicolumn{1}{c|}{1713.0}                                  & 0.07721                                  \\ \hline
MDME~\cite{li2022billion}                    & \multicolumn{1}{c|}{243.71} & \multicolumn{1}{c|}{0.04395}   & 0.8482 & \multicolumn{1}{c|}{1743.6}                                & 0.07843                                  \\ \hline
Production Model        & \multicolumn{1}{c|}{\underline{240.16}} & \multicolumn{1}{c|}{\underline{0.07436}}   & \underline{0.8629} & \multicolumn{1}{c|}{\underline{1710.5}}                                & \underline{0.10291}                                  \\ \hline
\textbf{OURS}                    & \multicolumn{1}{c|}{\textbf{238.50}*}  & \multicolumn{1}{c|}{\textbf{0.08708}*}   & \textbf{0.8693}* & \multicolumn{1}{c|}{\textbf{1698.5}*}                                & \textbf{0.12844}*                                  \\ \hline
Improv.                 & \multicolumn{1}{c|}{0.69\%} & \multicolumn{1}{c|}{17.11\%}   & 0.74\% & \multicolumn{1}{c|}{0.70\%}                                & 24.81\%                                  \\ \hline
\end{tabular}
\smallskip
\end{table}
To evaluate the effectiveness of our proposed collaborative-enhanced production model, we conducted experiments on a private dataset collected between August 1, 2022, and August 31, 2022. We divided the data into two sets: the first 30 days were used for training, and the data from the last day were reserved for testing. In addition to our production model, we compared it to other competitive models, including ZILN~\cite{arxiv2019ziln}, developed by Google, and MDME~\cite{li2022billion}, proposed by KuaiShou.
To provide a comprehensive performance comparison, we evaluated the models on two subsets of the data: all samples and samples with spending greater than zero. In the table below, our proposed model is denoted as "\textbf{OURS}" in Table~\ref{tab:over_offline}. The experiments are used to evaluate the performance of all models in predicting the spending money with RMSE and R2 Score~\cite{Wikipedia_2023_r2score}. AUC is used to evaluate whether the proposed model can predict whether the user downloaded the target game. 

From Table~\ref{tab:over_offline}, we can find our proposed model has surpassed all other models in all performance metrics, including the Linear Regression model, Multi-Layer Perceptron (MLP), Random Forest, XGBoost, Zero-Inflated Negative Binomial Regression (ZILN), Mixed Discrete and Continuous Output Model (MDME), as well as the existing production model.
Our proposed model achieved the best results on all samples and those with payments greater than zero, across Root Mean Square Error (RMSE), R2 Score, and AUC. Particularly in terms of the R2 Score, it showed an improvement of \textbf{17.11\%} for all samples and \textbf{24.81\%} for samples with payments greater than zero, compared to the production model. These findings undoubtedly illustrate that the collaborative-enhanced model we have designed can significantly enhance the performance of existing production models for spending money prediction tasks. 

\subsubsection{Online A/B Test}

\begin{figure}[]
    \begin{subfigure}{\columnwidth}
        \centering
        \includegraphics[width=0.9\columnwidth]{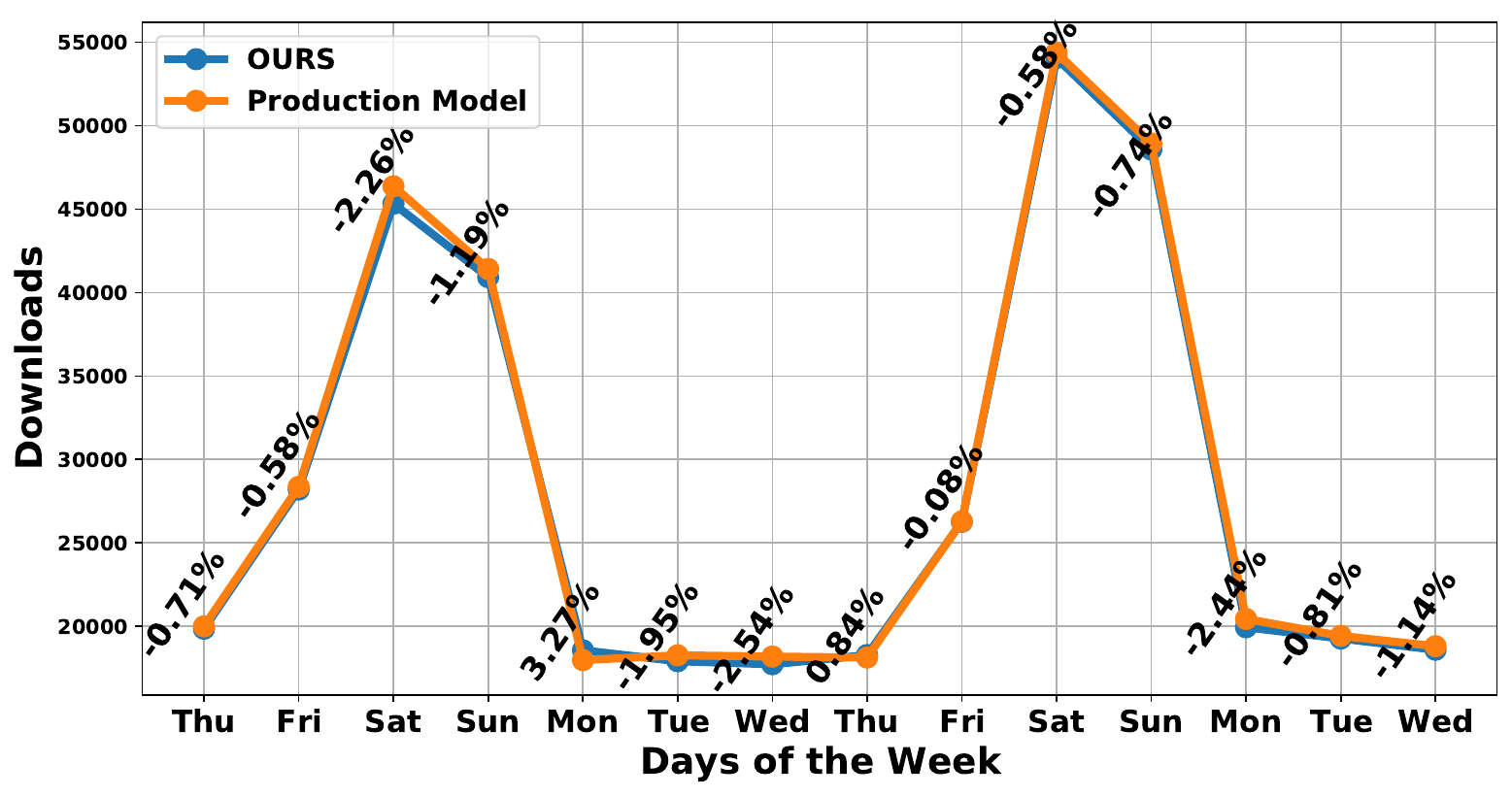}
        \caption{Download Data for the Date Range 20231019 to 20231101}
        \label{fig:download_data}
    \end{subfigure}

    \begin{subfigure}{\columnwidth}
        \centering
        \includegraphics[width=0.9\columnwidth]{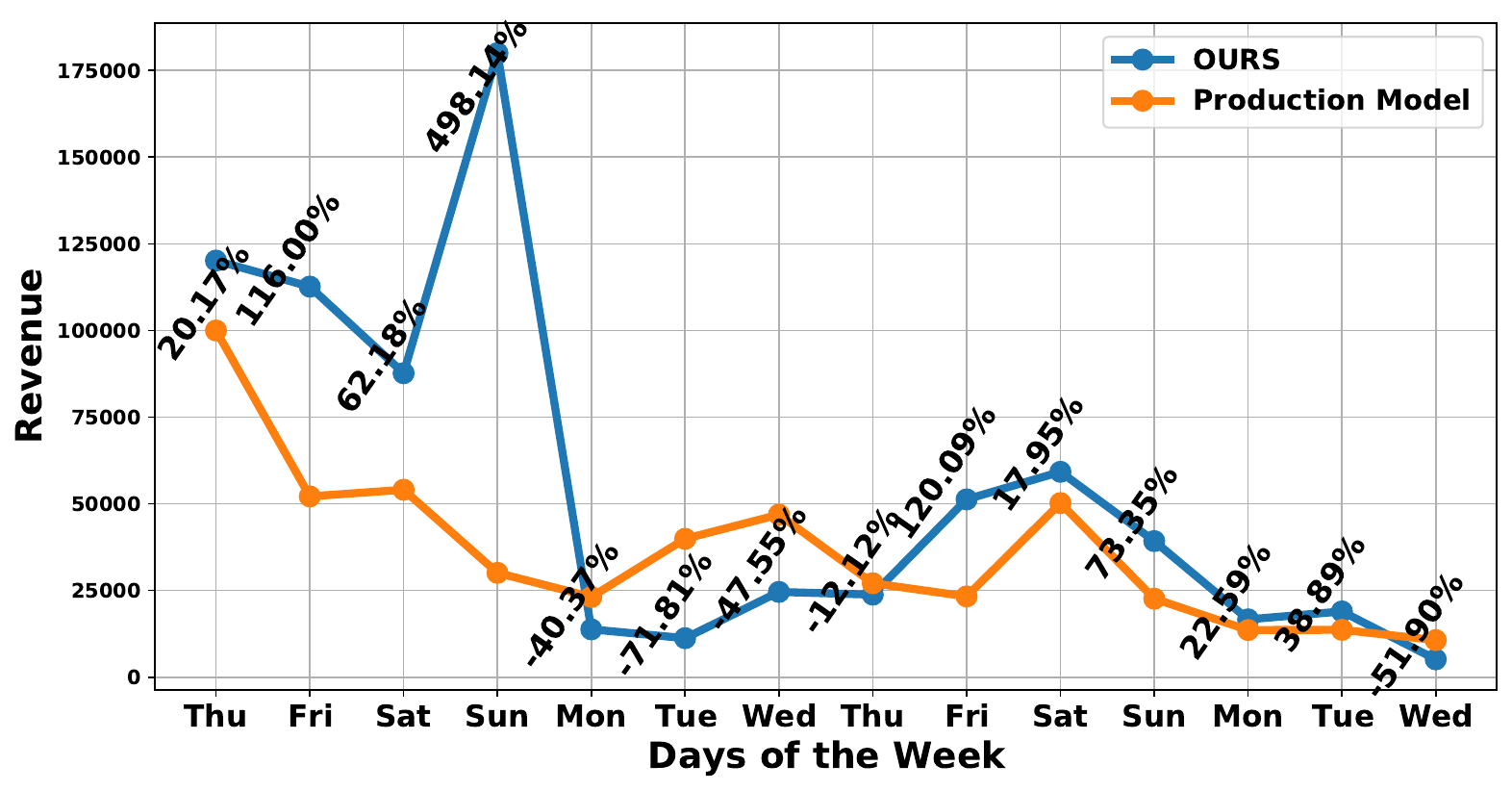}
        \caption{Revenue Data for the Date Range 20231019 to 20231101}
        \label{fig:revenue_data}
    \end{subfigure}
    
    \caption{User Game Downloads and Revenue on Newly Downloaded Games Comparison of Our Proposed Model and Production Model in an Online A/B Test from 2023-10-19 to 2023-11-01. The values corresponding to all data points are relative to protect the company's property. }
    \label{fig:down_revenue}
\end{figure}

An online A/B test was conducted from October 18, 2023 for a two-week observation period. The test was implemented in the Editor Recommendation Module on the Huawei Game Store, with 30\% of the recommended games downloads and experiment group. Our proposed collaborative-enhanced model was deployed in the experiment group to predict users' spending on newly downloaded games in the next 30 days. Based on this prediction, different games were recommended to different users. The production model served the control group. We focused on users' recommended games downloads and in-game purchases.

Results demonstrate that while the overall download rate decreased slightly by 0.88\%, the proposed model achieved a \textbf{50.65\%} improvement in revenue compared to the production model. The significant revenue gains, despite the minor decrease in downloads, suggest our model was able to recommend games that users were more likely to enjoy. This could reduce situations where users download more games due to unsatisfying recommendations, which explains the slight decrease in overall downloads.

Figure~\ref{fig:down_revenue} shows more detailed download and revenue patterns. Further analysis of the A/B test data indicates that the proposed model attained substantial revenue improvements consistently across weekdays and weekends. This points to the robustness of our collaborative approach across different user activity profiles.
Moreover, some weekly cycles were also observed. 
After obtaining confirmation from the business unit, we conducted a larger-scale test in the relevant recommendation scenarios and ultimately confirmed the formal implementation of this improvement into the respective product. 
As of February 2, 2024, the latest information we have obtained indicates that our proposed algorithm has achieved an \textbf{18.43}\% increase in revenue in all traffic tests in some recommendation scenarios.

\section{Related Work}

\subsection{Customer Lifetime Value Prediction}
In this paper, the research problem we study focuses on predicting users' spending on newly downloaded games~\cite{recsys2023cltv}. Forecasting spending is challenging due to uncertainties in timing and amount. However, LTV prediction has a long history of research since the 1920s. In 1920, M. Greenwood introduced RFM and NBD for disease and disaster prediction~\cite{greenwood1920inquiry}. In 1959, A.S.C. Ehrenberg applied RFM and NBD ideas to marketing~\cite{ehrenberg1959pattern}. In 1987, D.C. Schmittlein extended NBD to Pareto/NBD by mixing NBD with Gamma distribution of exponentials~\cite{schmittlein1987counting}.
Recent LTV studies can be categorized based on machine learning methods and whether a two-stage prediction is used. By ML methods, there are Bayesian approaches like NBD, Pareto/NBD, random forest-based methods~\cite{win2020randomforest}, and neural network models such as paper~\cite{sifa2018cltv}, TSUR~\cite{xing2021tsur}, perCLTV~\cite{zhao2023percltv}, and ExpLTV~\cite{zhang2023expltv}. By prediction stage, two-stage methods separately predict payers and payment amounts~\cite{vanderveld2016engagement, kdd2017cltv}, or payment frequencies and amounts per payment~\cite{fader2010customer, venkatesan2004customer}. Non-two-stage methods directly predict spending from user/product features and historical data~\cite{malthouse2005cltv, gupta2006modeling, sakia1992box}. However, most existing works predict future spending based on past transactions, which is inapplicable to our new game download scenario.
In summary, while LTV prediction has been studied for decades, forecasting user spending on newly downloaded games remains an open challenge, rendering most prior arts inapplicable. Developing suitable solutions for this problem requires exploring novel directions. We argue that one challenge of this task is the very oddly distributed user spending money label~\cite{valdivia2021customer}, which is also addressed in ZILN~\cite{arxiv2019ziln} and MDME~\cite{li2022billion}. In this paper, we argue collaborative signals can improve spending prediction, but label noise and imbalance in the dataset hinder model exploration. 

\subsection{Cross-Feature Models}
The user spending prediction task can be treated as a conversion rate (CVR) prediction problem. Hence, we can refer to cross-feature models used in CVR prediction. These models can automatically learn complex relationships between different features, such as FM~\cite{idcm2010fm}, Wide\&Deep~\cite{cheng2016widedeep}, DeepFM~\cite{guo2017deepfm}, xDeepFM~\cite{lian2018xdeepfm}, NFM~\cite{he2017nfm}, etc. Each model designs specific modules to enhance automatic feature engineering. Due to their strong modeling capabilities, DCN~\cite{wang2017dcn} and DCNV2~\cite{www2021dcnv2} have recently been considered state-of-the-art.
We employ in our production environment, we employ DCN, considering its time efficiency and good performance. Some other models like AutoInt~\cite{song2019autoint} and CAN~\cite{bian2022can} have also shown superior results, but their high time complexities prevent actual deployment. 
Cross-feature models enable efficient feature engineering by automatically capturing intricate feature interactions. The DCN family balances modeling power and time complexity, making it prevalent in real-world systems. 

\subsection{Collaborative-based Methods}
To infer users' potential spending on newly downloaded games without historical interaction records, collaborative filtering approaches are naturally promising solutions. Existing collaborative methods can be categorized into memory-based and model-based. Classical memory-based algorithms include item-based methods~\cite{sarwar2001item_based, ning2011slim}, while typical model-based approaches contain matrix factorization models like PMF~\cite{mnih2007pmf} and BPR~\cite{rendle2012bpr}. Some hybrid methods combine memory-based and model-based strategies, such as SVD++~\cite{kdd2008svd++} and NESCL~\cite{sun2023nescl}.
With the success of deep learning, neural collaborative filtering models like NeuMF~\cite{www2017neumf} have emerged. To address the sparsity issue, solutions like graph-based methods and harder negative sampling have been proposed~\cite{chen2020lrgccf, sigir2020lightgcn, mao2021ultragcn}. Self-supervised learning has been introduced to enhance recommender model training~\cite{khosla2020supcon, sigir2021sgl, sun2023nescl, le2020contrastive}. Other domains like social recommendation~\cite{sun2018arsr, wu2019diffnet, wu2020diffnet++} and news recommendation~\cite{an2019news_lstm} have also benefited from modeling collaborative signals. 
In addition, because the collected user interaction data contains noise and bias, some work has also been proposed to address such issues, such as ~\cite{wang2023ulc}. 
However, applying existing collaborative filtering approaches to our problem still faces some challenges. First, user IDs are unavailable, so new ways to model collaborative signals without IDs need to be explored. Second, full retraining on entire datasets is infeasible, so collaborative models that allow efficient incremental updates are preferred. Lastly, production systems already model some collaborative information via cross-features. Effectively combining explicit and implicit collaborative signals is non-trivial.
Although methods like Hi-Transformer~\cite{wu2021hitransformer} and NRMS~\cite{wu2019multi_head} learn user representations without IDs, they do not consider the dynamic update and signal combination issues. While CFFNN~\cite{yu2021cffnn} incorporates collaborative filtering and cross-features, it focuses less on integrating explicit and implicit signals. 
\section{Conclusion and Future Work}
In this work, we proposed a framework to enable stable training and evaluation of models predicting user spending on newly downloaded games. This addresses challenges from unpredictable user behavior and data characteristics like label variance and extremes. Within the framework, we developed a collaborative-enhanced model that represents user preferences and game features separately before concatenating them to predict spending.
The main contributions are: 1) The framework allows robust model exploration given uncertain user spending habits by standardizing spending data. 2) The collaborative model enhances prediction accuracy without relying on user IDs, enabling online training while protecting user privacy. And 3) experiments showed the collaborative model outperforms production models, increasing prediction accuracy by \textbf{17.11}\% on offline data and revenue by \textbf{50.65}\% in an online A/B test. Overall, the framework and collaborative modeling approach promise to maximize revenue from mobile games through more accurate user spending prediction, while ensuring model stability and user privacy. 
In the future, we aim to improve game value estimation in three aspects: enhancing model generalization, testing scalability with different datasets, and exploring new applications in streaming platforms and e-commerce.


\clearpage

\bibliographystyle{ACM-Reference-Format}
\bibliography{sample}

\end{document}